# Comments on the paper 'Growth and Characterization of Pure and Thiourea-Doped *L*-Alanine Single Crystals for NLO Devices'

*Bikshandarkoil R. Srinivasan[1] and  S. Natarajan[2*]*

[1]Department of Chemistry, Goa University, Goa, India 403004  srini@unigoa.ac.in
[2]School of Physics, Madurai Kamaraj University, Madurai, India 625021
[*]Corresponding author e-mail:    s_natarajan50@yahoo.com

**Abstract**

In this comment, we show that the pure and thiourea-doped glycyl-*L*-alanine hydrochloride (GLAH and TU-GLAH) crystals claimed to have been grown by the authors of the title paper (Journal of Russian Laser Research, **34**, 346-350 (2013)) are dubious crystals.

**Comment**

From the title of their paper [1], it appears that the authors describe the growth and characterization of pure and thiourea-doped *L*-alanine single crystals.  However, on reading the abstract it is learnt that the work described in the title paper is actually on the growth and characterization of pure and thiourea-doped glycyl-*L*-alanine hydrochloride (GLAH and TU-GLAH) crystals. According to the authors, these so called GLAH and TU-GLAH crystals were characterized by single crystal X-ray diffraction and were found to belong to the monoclinic space group $P2_12_12_1$ and the lattice parameters for pure GLAH were found to be in agreement with those of glycyl-*L*-alanine hydrochloride [2]. In this context, we wish to remark that not only the space group assignment but also the claim of the authors regarding the agreement of the cell parameters are incorrect as can be evidenced from Table 1. The authors did not report any CIF file to substantiate the space group and the structure.

A scrutiny of the experimental details in the title paper reveals that the authors have grown a so called glycyl-*L*-alanine hydrochloride GLAH crystal by slow evaporation of an aqueous solution containing stoichiometric amounts of glycine and *L*-alanine and excess of HCl following their earlier procedure.

In a recent paper [3] we have proved that no glycyl-*L*-alanine hydrochloride can be crystallized by mixing glycine and *L*-alanine in water with excess HCl as hydrochloric acid is not an appropriate reagent for peptide formation. We wish to mention that formation of glycyl-*L*-alanine (peptide) is a dehydration reaction and such a reaction cannot be performed in an aqueous medium. Our comments [3] are very much applicable for the growth of glycyl-*L*-alanine hydrochloride GLAH crystal described in the title paper to prove it is a dubious crystal. Since the GLAH crystal is questionable, the so called thiourea doped crystal TU-GLAH which is derived from it can also be declared as a dubious crystal. As the crystals grown by the authors of the title paper are dubious, other investigations such as second harmonic generation (SHG) study, microhardness and dielectric studies are meaningless and hence are not commented. In view of the above mentioned points, the claims made in the title paper are untenable and the paper is completely erroneous.

**Table 1** Unit cell comparison of authentic glycyl-*L*-alanine hydrochloride with so called GLAH and TU-GLAH crystals

| Crystal | $a$ / Å | $b$ / Å ($\beta$ / °) | $c$ / Å | Space Group | Ref |
|---|---|---|---|---|---|
| A so called glycyl-*L*-alanine hydrochloride GLAH | 5.73 | 18.27 (96°72") | 7.87 | $P2_12_12_1$[#] | [1] |
| A so called thiourea doped Glycyl-*L*-alanine hydrochloride monohydrate TU-GLAH | 5.63 | 18.15 (98°72") | 7.94 | $P2_12_12_1$[#] | [1] |
| Glycyl-*L*-alanine hydrochloride | 8.21(2) | 5.04(1) 100.1(2) | 10.82(2) | $P2_1$ | [2] |

[#] No CIF data reported; The space group is incompatible with the unit cell metrics.